# The Next Era of American Law
# Amid the Advent of Autonomous AI Legal Reasoning

**Dr. Lance B. Eliot**
Chief AI Scientist, Techbruim; Fellow, CodeX: Stanford Center for Legal Informatics
Stanford, California, USA

**Abstract**

Legal scholars have postulated that there have been three eras of American law to-date, consisting in chronological order of the initial Age of Discovery, the Age of Faith, and then the Age of Anxiety. An open question that has received erudite attention in legal studies is what the next era, the fourth era, might consist of, and for which various proposals exist including examples such as the Age of Consent, the Age of Information, etc. There is no consensus in the literature as yet on what the fourth era is, and nor whether the fourth era has already begun or will instead emerge in the future. This paper examines the potential era-elucidating impacts amid the advent of autonomous Artificial Intelligence Legal Reasoning (AILR), entailing whether such AILR will be an element of a fourth era or a driver of a fourth, fifth, or perhaps the sixth era of American law. Also, a set of meta-characteristics about the means of identifying a legal era changeover are introduced, along with an innovative discussion of the role entailing legal formalism versus legal realism in the emergence of the American law eras.

## 1 Background on Eras of American Law

In Section 1 of this paper, the eras of American law are introduced and addressed. Doing so establishes the groundwork for the subsequent sections. Section 2 introduces the Levels of Autonomy (LoA) of AI Legal Reasoning (AILR), which is instrumental in the discussions undertaken in Section 3. Section 3 provides an indication of the eras of American law as it applies to the LoA AILR. Section 4 provides various additional considerations and recommendations.

This paper then consists of these four sections:
- Section 1: Background on Eras of American Law
- Section 2: Autonomous Levels of AI Legal Reasoning
- Section 3: Next Eras and Autonomous AI Legal Reasoning
- Section 4: Additional Considerations and Future Research

### 1.1 Three Eras of American Law

The history of American law has been extensively studied and analyzed (see for example [29] [34] [60]). In the 1970s, legal scholar Grant Gilmore [32] proposed that the history of American law could be stratified into three distinct eras.

The first era was coined as the *Age of Discovery* and occurred from the 1800s until the Civil War, during which there was an initial formulation of a legal edifice for America. This was based to a great extent on the reuse of English common law, inexorably being shorn into a stylized and substantive instantiation that would become uniquely American law.

The second era lasted from the Civil War until WWI and was named as the *Age of Faith*. During this era, there was a purported attempt to perceive and shape the law as a form of rigorous science, out of which there were presumably legal truths that could axiomatically be discovered and derived. It was said to be a time when one ought to have utter faith that the law was right and just since it was essentially scientifically provable as such.

The third era arose following WWI and is referred to as the *Age of Anxiety*, and, when Gilmore wrote about these eras, he indicated that the Age of Anxiety was still underway. The anxiety being spurred as a realization that the faith in legal truths was mislaid and could no longer adequately serve as a foundational structure for understanding and maturation of the field of law.

As will be later discussed in this paper, there have been subsequent suggestions that we are either now in a fourth era or perhaps on the verge of such [7] [44], and thus it could be that we are already past the third era and into the fourth era. On the other hand, there does not appear to be any preponderance of agreement or consensus in the legal literature that a fourth era is indeed already taking place. Ergo, for purposes of this discussion herein, we assume that we are still in the third era and that the nature, timing, and impacts of a fourth era are not as yet collectively ascertained and nor cast into stone, as it were.

It is instructive to take a somewhat closer look at the considered prevailing three eras, doing so via the analyses and interpretations that have been made by legal scholars subsequent to Gilmore's groundbreaking book. On a related note, it can be said that the book was groundbreaking, though it is also notable to emphasize that Gilmore's work was based on the earlier efforts of Llewellyn [39]. Thus, it is significant and appropriate to saliently point out that Gilmore too indicates that his articulation of the three eras was based on the earlier work of Karl Llewellyn (from his respective book of 1960). Throughout the discussion herein, the primary reference for the three eras is based on Gilmore, though at the same time it is rightfully noteworthy to point out the earlier work by Llewellyn.

Consider the analysis by Webster [59] concerning Gilmore's eras. Per Webster: "The initial Age of Discovery, lasting from the early nineteenth century to the Civil War, is said to have been a golden age, likened perhaps somewhat excessively-to the period of late sixteenth century English theater or late eighteenth century Viennese music. It was an age when a society's best minds (for no particular reason) naturally gravitated towards a particular endeavor: the creation of an American legal system."

Especially well-known historical figures and legal jurists such as John Marshall, Joseph Story, and Lemuel Shaw are representative of this first era.

The second era was prompted by a slew of factors, as Webster [59] states: "As is often the case, golden ages flourish briefly and then disappear as new political, social, and economic realities overtake them. Such was the case with American law at about the time of the Civil War. The fluidity, innovation and imagination which had created the new system of American law were replaced by a far more formalistic system emphasizing stability, certainty, and predictability."

This shift towards greater formalism has been depicted as an assumption that the law could be cast as a type of science and therefore embody the corresponding rigors, as per Webster [59], some believed that the law was "capable of being synthesized or reduced to basic truths or principles which, like laws of physics or chemistry, could be used as the criteria for legal decision-making." This has been coined as the era of faith, meaning having a kind of righteousness faith in the law and believing without reservation that it was only a matter of scientifically discovering and codifying irrefutable and immutable laws.

Bobbitt [7] points out that the "Age of Faith lasted from the Civil War through World War I, and was notable for the Olympian status it accorded law and its demigods, including Christopher Columbus Langdell and Oliver Wendell Holmes, Jr. Langdell believed that law was a science from which scientific truths could be derived, and even the skeptical Holmes, according to Gilmore, refined and judicialized Langdellianism." Webster equally points out that the law as science was replaced or desired to be replaced with an indication that law is a social science. There is some ongoing debate about whether the third era is therefore a complete rejection and departure of the second era, or perhaps a refinement, demonstrably so, and a shift toward a greater realism about the law.

This is how Gilmore expresses the semblance of the third era [32]:

> "As lawyers we will do well to be on our guard against any suggestion that, through law, our society can be reformed, purified or saved. The function of law, in a society like our own, is altogether more modest and less apocalyptic. It is to provide a mechanism for the settlement of disputes in the light of broadly conceived principles on whose soundness, it must be assumed, there is a general consensus among us. If the assumption is wrong, if there is no consensus, then we are headed for war, civil strife, and revolution, and the orderly administration of justice will become an irrelevant, nostalgic whimsy until the social fabric has been stitched together again and a new consensus has emerged."

Subsequent to Gilmore's book there have been extensive and ongoing deliberations about the naming used to reflect the three eras.

As pointed out by Webster [59]: "His book runs the risk of being criticized for affixing catchwords to time periods for which other and perhaps better names have already been given-liberalism, activism, and progressivism are just a few." In whatever manner the eras might be so coined, there seems little doubt about the importance of pointing out the eras per se and engaging in an overarching discussion about the evolution of America law, per Bobbitt [7]: "Gilmore's lectures satisfied and mesmerized their audience, and they were soon fashioned into a book, also titled *The Ages of American Law*, which became a foundational text for introducing law students to American legal studies."

In brief, the three eras as postulated seem to have become relatively accepted by the legal literature. Undoubtedly, arguments can be made as to whether these are the "right" eras and that perhaps the American law timeline period can be sliced in a different fashion. For example, suppose there have been ten eras, brandishing this notion as a provocative point of conjecture, or that there has only been one era and it has lasted since the beginnings of American law. Such provocateur outside-the-box considerations are not in the scope of this paper and it is taken at face value that there have been three eras and those three eras are reasonably stated and depicted by Gilmore and others that have similarly allied with the generally accepted stratification.

**1.1 Important Influence of Legal Realism**

Another notable perspective on the three eras is that they are all interconnected by an at-times hidden underlying variable. The postulated crux of the three eras has to do with the philosophical milieu known as legal formalism, which, in short, legal formalism can be defined as [20]: "A theory that legal rules stand separate from other social and political institutions. According to this theory, once lawmakers produce rules, judges apply them to the facts of a case without regard to social interests and public policy." In contrast, there is legal realism, defined as [20]: "A theory that all law derives from prevailing social interests and public policy. According to this theory, judges consider not only abstract rules, but also social interests and public policy when deciding a case."

In the prevailing implications of the three eras, the first era was a mild form of legal formalism but that was grappling with establishing the new American law and thus only modestly embraced such formalism precepts, while the second era was a dramatic extension and gravitation toward legal formalism, and then the third era a kind of retreat from legal formalism in place of alternatively embracing legal realism.

Per Webster [59]: "Had those early legal figures instead adopted a formalistic approach to the problem of creating a new law for America, they would have sought to adopt the far more static and fixed principles of English common law and the English legal system. To an extent, of course, this happened. But for reasons relating, in part, to political and social hostility towards the utilization of an English system, English common law and, more importantly, the English approach to law were not simply imported to these shores without question. Because those early lawyers were not formalists, an innovative, creative and uniquely American system arose." And, as already cited in the prior subsection [59], the second era was a flourishing of legal formalism: "The fluidity, innovation and imagination which had created the new system of American law were replaced by a far more formalistic system emphasizing stability, certainty, and predictability."

And then, the third era, as described accordingly by Bobbitt [7]: "After that came the Age of Anxiety, Gilmore's own era, an Age when legal realism gnawed through the core assumptions of the Age of Faith and the nation groped unsuccessfully for new creeds to replace them."

The legal profession and legal scholars continue to debate the merits of legal formalism versus legal realism. Perhaps one of the most infamous euphemisms about the law is that supposedly a good lawyer knows the law, while a great lawyer knows the judge. In this paper, we do not address this at times acrimonious discourse and simply acknowledge that these matters are still being examined and assessed by ongoing legal research. In any case, if one is seeking a measure or barometer toward identifying the pattern underlying the eras, it could be reasonably argued that the degree of legal formalism has been crucial if not a predominant factor at play. This will be further explored in Section 3.

The next section of this paper introduces the autonomous levels of AI Legal Reasoning, doing so to then aid Section 3. Section 3 explores how the eras of American law and the elucidations of the fourth era can be explored in light of the autonomous levels of AI Legal Reasoning. Section 4 provides some conclusionary remarks and also an indication of recommended future research.

## 2 Autonomous Levels of AI Legal Reasoning

In this section, a framework for the autonomous levels of AI Legal Reasoning is summarized and is based on the research described in detail in Eliot [20].

These autonomous levels will be portrayed in a grid that aligns with key elements of autonomy and as matched to AI Legal Reasoning. Providing this context will be useful to the later sections of this paper and will be utilized accordingly.

The autonomous levels of AI Legal Reasoning are as follows:

Level 0: No Automation for AI Legal Reasoning

Level 1: Simple Assistance Automation for AI Legal Reasoning

Level 2: Advanced Assistance Automation for AI Legal Reasoning

Level 3: Semi-Autonomous Automation for AI Legal Reasoning

Level 4: Domain Autonomous for AI Legal Reasoning

Level 5: Fully Autonomous for AI Legal Reasoning

Level 6: Superhuman Autonomous for AI Legal Reasoning

### 2.1 Details of the LoA AILR

See **Figure A-1** for an overview chart showcasing the autonomous levels of AI Legal Reasoning as via columns denoting each of the respective levels.

See **Figure A-2** for an overview chart similar to Figure A-1 which alternatively is indicative of the autonomous levels of AI Legal Reasoning via the rows as depicting the respective levels (this is simply a reformatting of Figure A-1, doing so to aid in illuminating this variant perspective, but does not introduce any new facets or alterations from the contents as already shown in Figure A-1).

#### 2.1.1 Level 0: No Automation for AI Legal Reasoning

Level 0 is considered the no automation level. Legal reasoning is carried out via manual methods and principally occurs via paper-based methods.

This level is allowed some leeway in that the use of say a simple handheld calculator or perhaps the use of a fax machine could be allowed or included within this Level 0, though strictly speaking it could be said that any form whatsoever of automation is to be excluded from this level.

#### 2.1.2 Level 1: Simple Assistance Automation for AI Legal Reasoning

Level 1 consists of simple assistance automation for AI legal reasoning.

Examples of this category encompassing simple automation would include the use of everyday computer-based word processing, the use of everyday computer-based spreadsheets, the access to online legal documents that are stored and retrieved electronically, and so on.

By-and-large, today's use of computers for legal activities is predominantly within Level 1. It is assumed and expected that over time, the pervasiveness of automation will continue to deepen and widen, and eventually lead to legal activities being supported and within Level 2, rather than Level 1.

### 2.1.3 Level 2: Advanced Assistance Automation for AI Legal Reasoning

Level 2 consists of advanced assistance automation for AI legal reasoning.

Examples of this notion encompassing advanced automation would include the use of query-style Natural Language Processing (NLP), Machine Learning (ML) for case predictions, and so on.

Gradually, over time, it is expected that computer-based systems for legal activities will increasingly make use of advanced automation. Law industry technology that was once at a Level 1 will likely be refined, upgraded, or expanded to include advanced capabilities, and thus be reclassified into Level 2.

### 2.1.4 Level 3: Semi-Autonomous Automation for AI Legal Reasoning

Level 3 consists of semi-autonomous automation for AI legal reasoning.

Examples of this notion encompassing semi-autonomous automation would include the use of Knowledge-Based Systems (KBS) for legal reasoning, the use of Machine Learning and Deep Learning (ML/DL) for legal reasoning, and so on.

Today, such automation tends to exist in research efforts or prototypes and pilot systems, along with some commercial legal technology that has been infusing these capabilities too.

### 2.1.5 Level 4: Domain Autonomous for AI Legal Reasoning

Level 4 consists of domain autonomous computer-based systems for AI legal reasoning.

This level reuses the conceptual notion of Operational Design Domains (ODDs) as utilized in the autonomous vehicles and self-driving cars levels of autonomy, though in this use case it is being applied to the legal domain [15] [17] [18]. Essentially, this entails any AI legal reasoning capacities that can operate autonomously, entirely so, but that is only able to do so in some limited or constrained legal domain.

### 2.1.6 Level 5: Fully Autonomous for AI Legal Reasoning

Level 5 consists of fully autonomous computer-based systems for AI legal reasoning.

In a sense, Level 5 is the superset of Level 4 in terms of encompassing all possible domains as per however so defined ultimately for Level 4. The only constraint, as it were, consists of the facet that the Level 4 and Level 5 are concerning human intelligence and the capacities thereof. This is an important emphasis due to attempting to distinguish Level 5 from Level 6 (as will be discussed in the next subsection)

It is conceivable that someday there might be a fully autonomous AI legal reasoning capability, one that encompasses all of the law in all foreseeable ways, though this is quite a tall order and remains quite aspirational without a clear cut path of how this might one day be achieved. Nonetheless, it seems to be within the extended realm of possibilities, which is worthwhile to mention in relative terms to Level 6.

### 2.1.7 Level 6: Superhuman Autonomous for AI Legal Reasoning

Level 6 consists of superhuman autonomous computer-based systems for AI legal reasoning.

In a sense, Level 6 is the entirety of Level 5 and adds something beyond that in a manner that is currently ill-defined and perhaps (some would argue) as yet unknowable. The notion is that AI might ultimately exceed human intelligence, rising to become superhuman, and if so, we do not yet have any viable indication of what that superhuman intelligence consists of and nor what kind of thinking it would somehow be able to undertake.

Whether a Level 6 is ever attainable is reliant upon whether superhuman AI is ever attainable, and thus, at this time, this stands as a placeholder for that which might never occur. In any case, having such a placeholder provides a semblance of completeness, doing so without necessarily legitimatizing that superhuman AI is going to be achieved or not. No such claim or dispute is undertaken within this framework.

# 3 Next Eras and Autonomous AI Legal Reasoning

As outlined in Section 1, make the assumption that there are three eras of American law and that they have been appropriately and indubitably identified and correctly typified.

See **Figure B-1** for a visual illustration of the three eras.

The logical questions that naturally flow from those three presumed eras consists of:
- What will be the fourth era?
- When will the fourth era begin (or has it already)?
- What is the basis for asserting there will be a fourth era?
- How will the fourth era be differentiated from the prior eras?
- To what degree does legal formalism partake in a fourth era?
- And so on.

There have been various proffered proposals about a fourth era. None of the postulated fourth eras has seemed to gain traction, as yet. This would appear to leave open for the time being the possibility of considering the nature and significance of a fourth era. Presumably, we are either not in a fourth era, or we might be in a fourth era and are generally unaware that we are.

For those that might believe it folly or valueless to speculate about a fourth era, this kind of matter is actually of both a notable theoretical and practical significance. By being able to anticipate the fourth era, we might collectively as a society and especially within the legal field be able to prepare accordingly for what is to come, along with the added potential of shaping or altering course if the emergent fourth era seems untoward or otherwise undesirable. For legal practitioners, knowing what the fourth era constitutes could aid significantly in their training and attention, and be a crucial harbinger of what the practice of law is coming to possibly become.

Consider two especially noted propositions for the possible fourth era, which are respectively known as the Age of Consent [7] and the other referred to as the Age of Information [44]. Both were proposed at approximately the same time and of relatively recent note (doing so around the year 2015).

A proposed fourth era coined as the Age of Consent was postulated by Bobbitt [7]: "As we enter the Age of Consent, the era of a new, already emerging constitutional order that puts the maximization of individual choice at the pinnacle of public policy, it would be well to appreciate the structuring role for choice that American law has always provided. Far from obviating the need for our consciences, our laws structure a necessary role for them. That highly structured role is reflected in representative government (rather than plebiscites), in the composition of juries (rather than mobs, even when they form over the Internet rather than outside a jail), in the belief in liberal education (rather than indoctrination), in the responsibility of judges and lawyers to shape as well as defend the Constitution that gives them unique power. Those structures will be strictly scrutinized in this era, as they should be. How else will these habits and practices find defenders unless they are convinced, after rigorous examination, that this way of structuring choice is worthy of defense?"

Another proposed fourth era is called the Age of Information, postulated by McGinnis and Wasick [44]: "Today we inhabit the Age of Information and this age is creating a new synthesis for the structure of law. If the Age of Faith required formalism to regulate the legal world, the Age of Information, like the Age of Anxiety, accepts that many factors may influence the law. But the Age of Information, like the Age of Faith, has greater confidence in creating legal clarity. Both the Age of Information and the Age of Faith have their gods of legal order, but if the god of the Age of Faith was formalism, today's god is computable realism. The rise of computable standards and dynamic rules will be this age's contribution to legal expression."

See **Figure B-2** for a visual illustration indicative of the fourth eras postulated.

An intriguing and potentially co-existent matter entails the advent of autonomous AI Legal Reasoning (AILR), as outlined generally in the prior section of this paper. Consider whether the emergence of autonomous AILR will be an element of the fourth era, or whether it is conceivable that autonomous AILR would be more than merely an ingredient and essentially constitute the namesake of a fourth era.

In the Age of Information, there is ample indication of the importance of AI in the law and how it will be a considered element. The Age of Consent does not directly tie the role of AI and the law into its depiction, though it is possible to discern underlying aspects that could be construed within the framework indicated. An additional argument to be made is that perhaps, if we are already in a fourth era, it could be that the rise of autonomous AILR might be a noteworthy contributor to the fourth era and ultimately be the forerunner or instigator of a fifth era.

See **Figure B-3** for a visual illustration of these possibilities.

Aligning the Section 2 indication of the levels of autonomy for AI Legal Reasoning, it is instructive to consider the impacts per each of the levels thereof.

See **Figure B-4** for an indication of the LoA AILR with the added indication accordingly.

In brief, the levels and the alignment to the eras appear to consist of:

- Level 0: In Eras 1, 2, 3+
- Level 1: In Eras 3, (4+)
- Level 2: In Eras 3, (4+)
- Level 3: In Eras 3, {4+)
- Level 4: Shape Eras 4 or 5
- Level 5: Define Eras 4 or 5
- Level 6: Be Eras 5 or 6

Here is an overview explanation associated with these stated possibilities.

At Level 0, the lack of computer-based automation was certainly the case during the first era, and during the second era too. During the third era, there are still ongoing examples of the lack of computer-based automation being utilized. It is presumably the case that the law will always be amenable to the nonuse of automation, thus the indication of "3+" meaning from the third era onward. An outstretched case can be made that there will be a future era during which computer-based automation will be so pervasive and so crucial that the law cannot exist or be undertaken without it, but this seems an outlier consideration at this time.

At Level 1, the advent of simple computer-based automation emerged in the third era and continues to this day. Presumably, this will continue into the fourth era and beyond. It would though seem that Level 1 is substantively lacking in the sense that it would not drive or materially spark a shift into a fourth era and thus the "(4+)" is used to indicate as such.

At Level 2, the advent of advanced computer-based automation also emerged in the third era and continues to this day. Presumably, this will continue into the fourth era and beyond. It would though seem that Level 2 is substantively lacking in the sense that it would not drive or materially spark a shift into a fourth era and thus the "(4+)" is used to indicate as such.

At Level 3, the advent of semi-autonomous AILR automation is gradually appearing in the prevailing third era and continues to this day. Presumably, this will continue into the fourth era and beyond. It would though seem that Level 3 is substantively lacking in the sense that it would not drive or materially spark a shift into a fourth era, despite whatever otherwise innovative prototypes or tryouts are launched, and thus the "(4+)" is used to indicate as such.

If Level 4 can be achieved, this would seem to be a notable basis for shaping or possibly even instigating the next era, whether it be the fourth era or a fifth era.

If Level 5 can be achieved, this would seem to be of such a transformative facet that it would dramatically define a fourth or fifth era.

If Level 6 can be achieved, this would be an even more momentous transformative facet, perhaps driving a fifth or sixth era.

### 3.1 Meta-Characteristics of Eras

This subsection introduces a strawman set of meta-characteristics that might be used to aid in delineating and identify the eras.

These five core facets are proffered:

- Distinctive
- Substantive
- Observable
- Paradigmatic
- Explainable

See **Figure B-5** for a visual illustration of these core facets.

Consider a brief explanation for each of the meta-characteristics, covering each of the key foundations for the application of each one:

- Distinctiveness entails that an era, regardless of sequence and whether already existent or still inexistent, showcases boundaries that can be distinctly identified, allowing us to realize that the era exists. Without distinctiveness, an alleged era would be indistinguishable presumably from the prior era, being no more than a blur or presumed extension of that era, and therefore could not demonstrably and verifiably be claimed as an era.

- Substantive refers to the era being of a convincing magnitude and substance to warrant the era moniker. If a proposed era was distinctive but without sufficient import or degree, it would be readily arguable whether it truly standalone as an era or might be a momentary aberration in the existing prevailing era.

- Observable refers to the era being detectible, such that if there is no means to discern that the era exists, it would be presumably imaginary and thusly suspect to being counted as real or notable.

- Paradigmatic refers to the aspect that an era would presumably need to fit within the paradigm associated with what eras are contended to be composed of. This is not to suggest that an era might breakaway from the existent paradigm, which certainly could be the case, though generally it is assumed that the collectively accepted paradigm is sufficient (obviously, a new paradigm is always worthy of consideration).

- Finally, it is suggested that an era would need to be explainable, which involves being able to provide a rationale and explication of why an era is thought to exist. This is not an especially necessary condition, since it is conceivable that an era exists and has resisted being naturally explained, but nonetheless it can be argued that without a sufficient explanation for the era it is unlikely to gain consensus for acknowledging its existence and substance.

The meta-characteristics are not numbered since to do so might imply prioritization or ranking. They are each of their own merit. In addition, they are a collective set, without which any individual one would be less capable and certainly lacking in robustness to indicate the entire arch or form of an era.

**3.2 Legal Formalism As Era Triggering Mechanism**

In this subsection, further discussion about the role of legal formalism is undertaken.

As mentioned in Section 1, there are assertions that legal formalism is at the underpinning of the three eras. See **Figure B-6** for a visual illustration of the legal formalism underpinning facets.

Consider this proffered explanation of the legal formalism underlaying role in the eras matters:

- It has been speculated that legal formalism began with an initial and yet somewhat distracted driving force in the making of Era 1, and thus on a spectrum of legal formalism ranging from none to some maximal amount, there is an arrow shown in Figure B-6 for Era 1 indicating this forward movement in the direction towards greater adoption of legal formalism during the first era.

- In the second era, it has been speculated in the literature that legal formalism took an even greater dominance, and thus in Figure B-6 for Era 2 there is an additional arrow further extending the legal formalism on the spectrum provided.

- For the third era, it has been speculated that legal formalism encountered retreatment, giving way to legal realism. As such, the arrow of legal formalism for Era 3 in Figure B-6 is shown beneath the earlier two and indicative of a lessening on the legal formalism spectrum.

- Regarding a fourth era, if the breakpoints or triggering mechanism that demarks the eras is indeed the legal formalism factor, presumably a divergent rise or lessening of legal formalism might be an earmark for the fourth era. This is not to assert that legal formalism is a cause-and-effect of the eras, which it might be, or it might be an indicator that correlates thereto, and thus this should be interpreted cautiously. In any case, if there was a substantive boost to legal formalism, perhaps this would be a sign or signal of a revival and might necessitate ascertaining whether a new era has emerged (see Figure B-6, Era 4). Likewise, if there was a substantive lessening of legal formalism, perhaps this would be a sign or signal of the obviating of legal formalism and might necessitate ascertaining whether a new era has emerged (see Figure B-6, Era 4).

Note that this depiction is merely an invigorating means to conceptualize the American law era shifts, and not intended to be a prescriptive or otherwise determinative indicator regarding the eras.

**4 Additional Considerations and Future Research**

As earlier indicated, legal scholars have postulated that three eras of American law have occurred to-date, consisting in chronological order of the initial Age of Discovery, the Age of Faith, and then the Age of Anxiety. Though there appears to be substantive consensus and acceptance of the three eras, there is the possibility that some other means of stratifying the history of American law could provide a new set of eras.

In any case, assuming that the three eras are the prevailing viewpoint, the open question of what is the fourth era remains available for ongoing discussion and debate. There is no consensus in the literature as yet on what the fourth era is, and nor whether the fourth era has already begun or will instead emerge in the future. This paper has examined the potential impacts due to the advent of autonomous Artificial Intelligence Legal Reasoning (AILR) on the question of the next era, including whether such AILR will be an element of a fourth era or a driver of a fourth, fifth, or perhaps sixth era of American law.

Also, a set of meta-characteristics about the means of identifying a legal era changeover have been introduced, along with an innovative discussion of the role entailing legal formalism versus legal realism in the emergence of the American law eras.

Future research is needed to explore in greater detail the manner and means by which AI-enablement will occur in the law along with the potential for both positive and adverse consequences. Autonomous AILR is likely to materially impact the eras of American law, including as a minimum playing a notable or potentially pivotal role in the next era(s), and having the possibility of shaping and instigating future eras altogether.

**About the Author**

Dr. Lance Eliot is the Chief AI Scientist at Techbrium Inc. and a Stanford Fellow at Stanford University in the CodeX: Center for Legal Informatics. He previously was a professor at the University of Southern California (USC) where he headed a multi-disciplinary and pioneering AI research lab. Dr. Eliot is globally recognized for his expertise in AI and is the author of highly ranked AI books and columns.

**Figure A-1**

| Level | Descriptor | Examples | Automation | Status |
|---|---|---|---|---|
| | | **AI & Law: Levels of Autonomy For AI Legal Reasoning (AILR)** | | |
| 0 | No Automation | Manual, paper-based (no automation) | None | De Facto - In Use |
| 1 | Simple Assistance Automation | Word Processing, XLS, online legal docs, etc. | Legal Assist | Widely In Use |
| 2 | Advanced Assistance Automation | Query-style NLP, ML for case prediction, etc. | Legal Assist | Some In Use |
| 3 | Semi-Autonomous Automation | KBS & ML/DL for legal reasoning & analysis, etc. | Legal Assist | Primarily Prototypes & Research Based |
| 4 | AILR Domain Autonomous | Versed only in a specific legal domain | Legal Advisor (law fluent) | None As Yet |
| 5 | AILR Fully Autonomous | Versatile within and across all legal domains | Legal Advisor (law fluent) | None As Yet |
| 6 | AILR Superhuman Autonomous | Exceeds human-based legal reasoning | Supra Legal Advisor | Indeterminate |

*Figure 1: AI & Law - Autonomous Levels by Rows*     *Source Author: Dr. Lance B. Eliot*     V1.3

**Figure A-2**

### AI & Law: Levels of Autonomy For AI Legal Reasoning (AILR)

| | Level 0 | Level 1 | Level 2 | Level 3 | Level 4 | Level 5 | Level 6 |
|---|---|---|---|---|---|---|---|
| **Descriptor** | No Automation | Simple Assistance Automation | Advanced Assistance Automation | Semi-Autonomous Automation | AILR Domain Autonomous | AILR Fully Autonomous | AILR Superhuman Autonomous |
| **Examples** | Manual, paper-based (no automation) | Word Processing, XLS, online legal docs, etc. | Query-style NLP, ML for case prediction, etc. | KBS & ML/DL for legal reasoning & analysis, etc. | Versed only in a specific legal domain | Versatile within and across all legal domains | Exceeds human-based legal reasoning |
| **Automation** | None | Legal Assist | Legal Assist | Legal Assist | Legal Advisor (law fluent) | Legal Advisor (law fluent) | Supra Legal Advisor |
| **Status** | De Facto – In Use | Widely In Use | Some In Use | Primarily Prototypes & Research-based | None As Yet | None As Yet | Indeterminate |

*Figure 2: AI & Law - Autonomous Levels by Columns*  *Source Author: Dr. Lance B. Eliot*

V1.3

**Figure B-1**

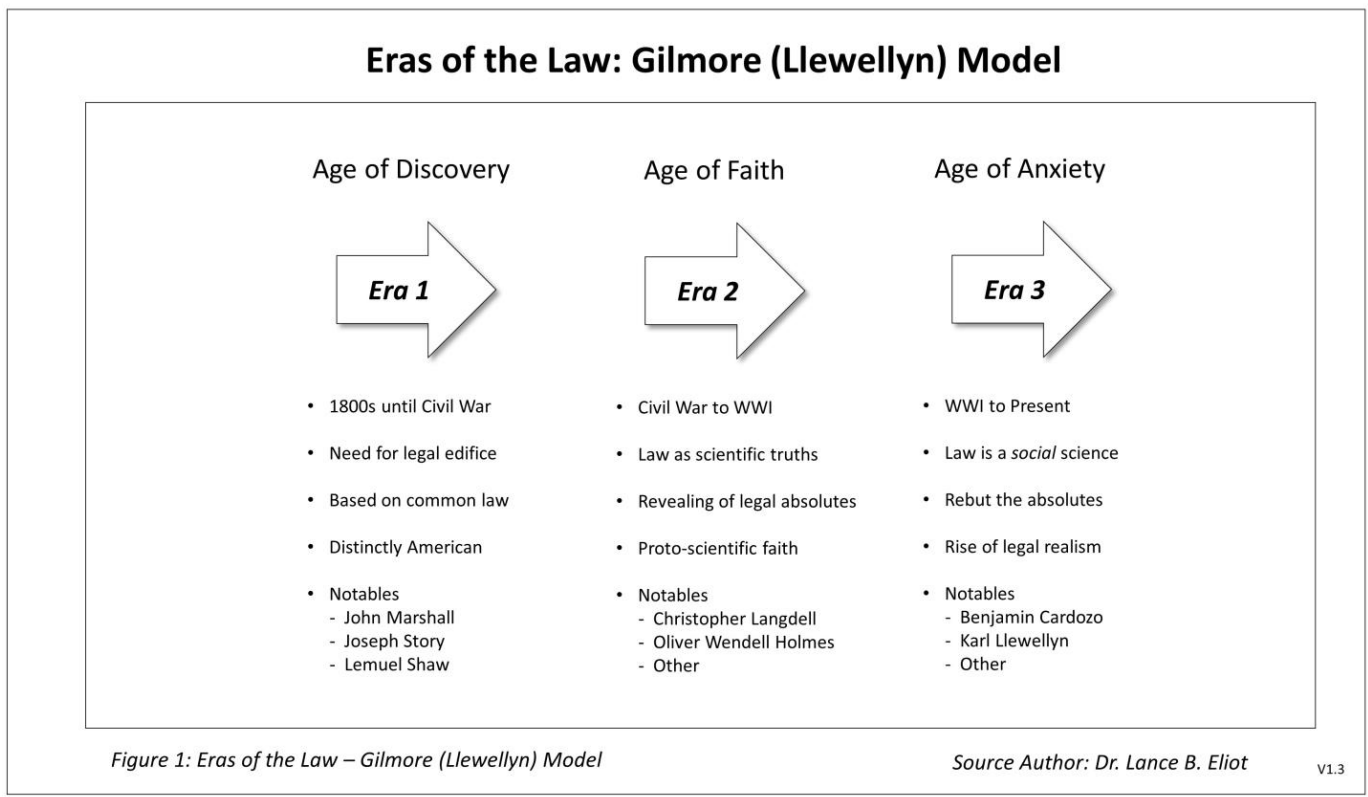

Figure 1: Eras of the Law – Gilmore (Llewellyn) Model

**Figure B-2**

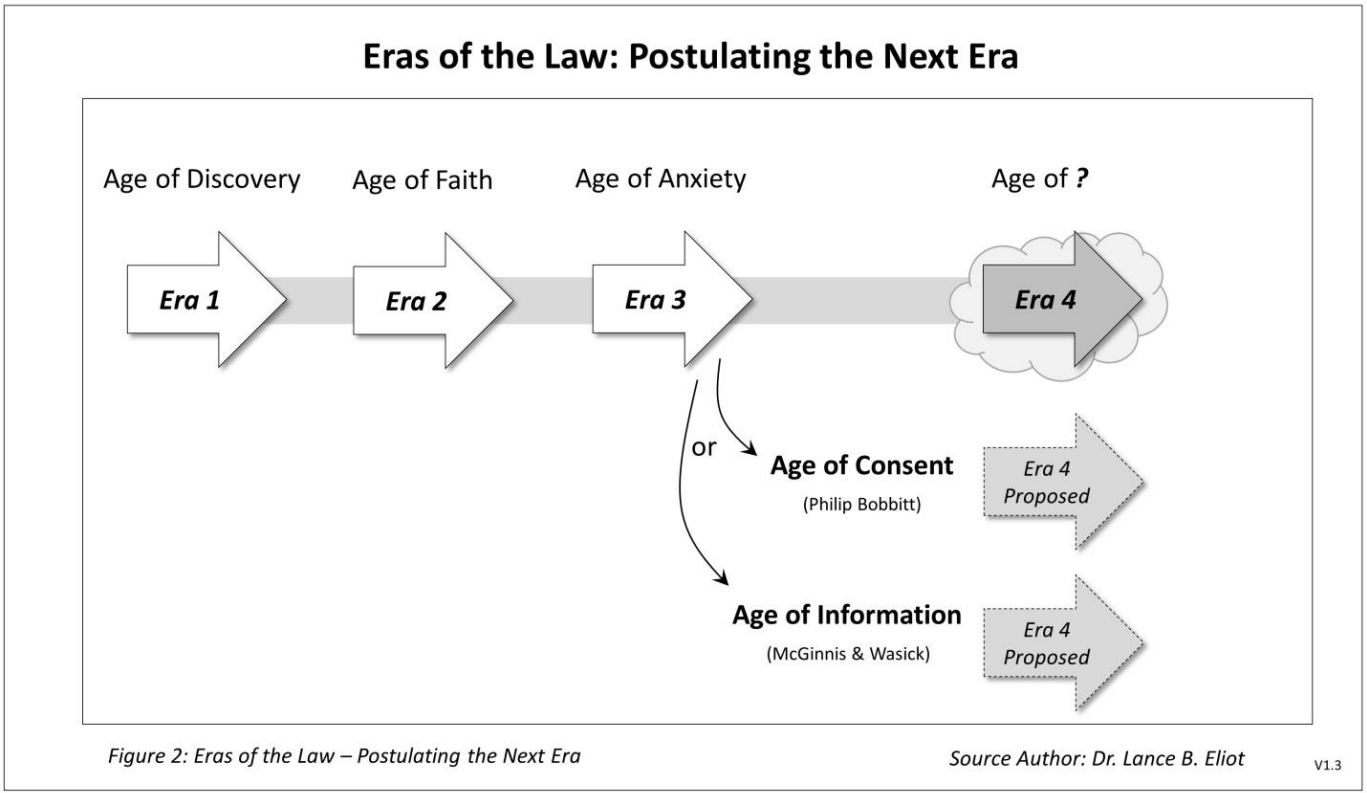

**Figure B-3**

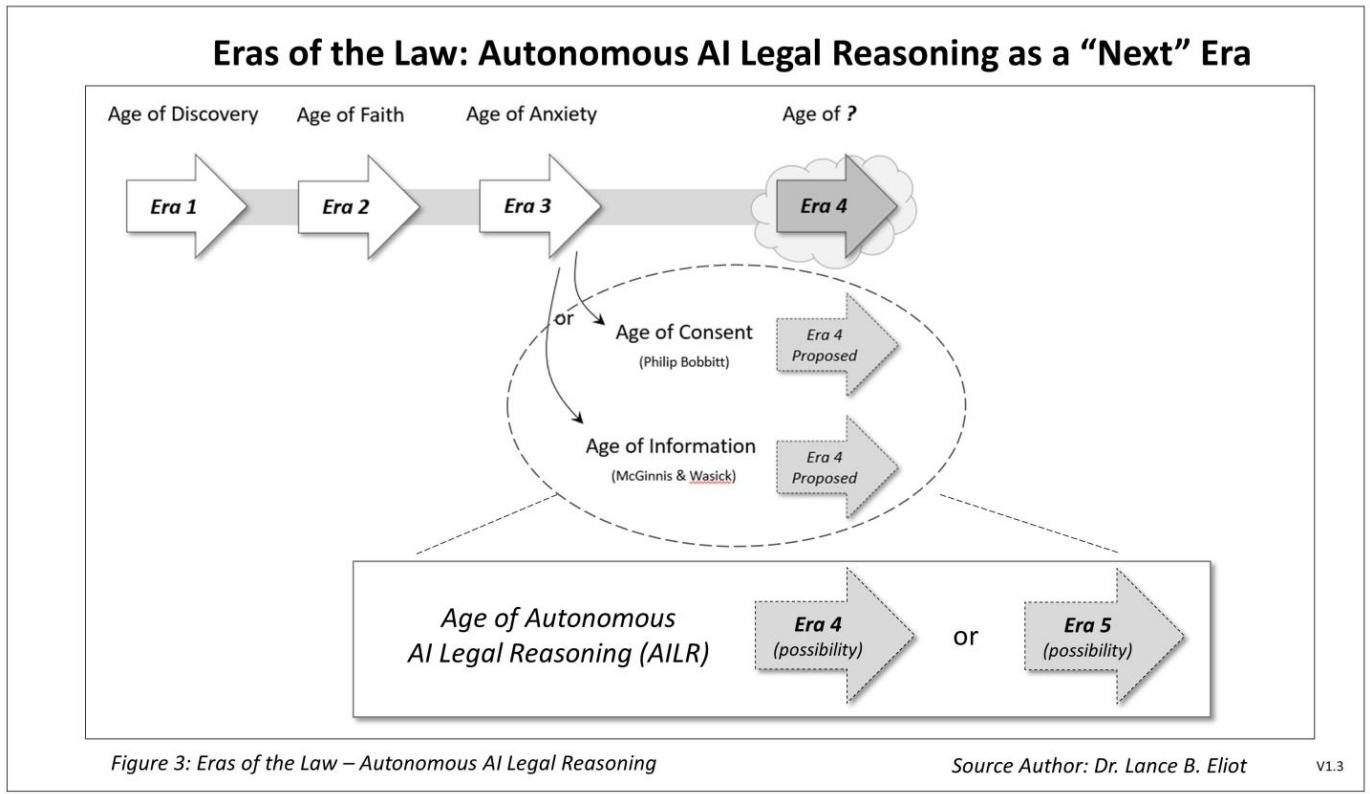

**Figure B-4**

### Eras of the Law: Levels of Autonomy For AI Legal Reasoning (AILR)

| | Level 0 | Level 1 | Level 2 | Level 3 | Level 4 | Level 5 | Level 6 |
|---|---|---|---|---|---|---|---|
| **Descriptor** | No Automation | Simple Assistance Automation | Advanced Assistance Automation | Semi-Autonomous Automation | AILR Domain Autonomous | AILR Fully Autonomous | AILR Superhuman Autonomous |
| **Examples** | Manual, paper-based (no automation) | Word Processing, XLS, online legal docs, etc. | Query-style NLP, ML for case prediction, etc. | KBS & ML/DL for legal reasoning & analysis, etc. | Versed only in a specific legal domain | Versatile within and across all legal domains | Exceeds human-based legal reasoning |
| **Automation** | None | Legal Assist | Legal Assist | Legal Assist | Legal Advisor (law fluent) | Legal Advisor (law fluent) | Supra Legal Advisor |
| **Status** | De Facto – In Use | Widely In Use | Some In Use | Primarily Prototypes & Research-based | None As Yet | None As Yet | Indeterminate |
| **Eras of the Law by Levels of Autonomy** | In Eras 1, 2, 3+ | In Eras 3, (4+) | In Eras 3, (4+) | In Eras 3, (4+) | Shape Eras 4 or 5 | Define Eras 4 or 5 | Be Eras 5 or 6 |

*Figure 4: Eras of the Law - Autonomous Levels of AILR by Columns*     *Source Author: Dr. Lance B. Eliot*     V1.3

**Figure B-5**

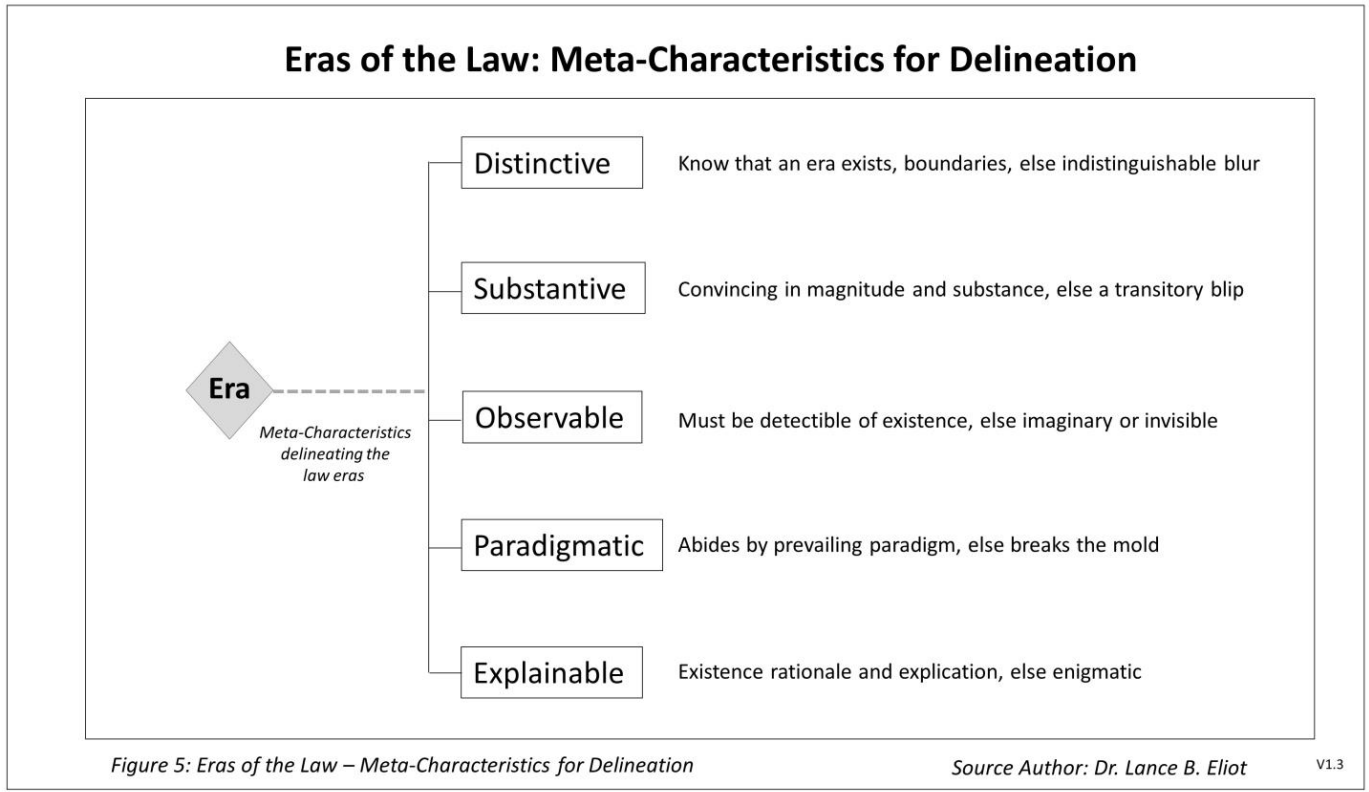

**Figure B-6**

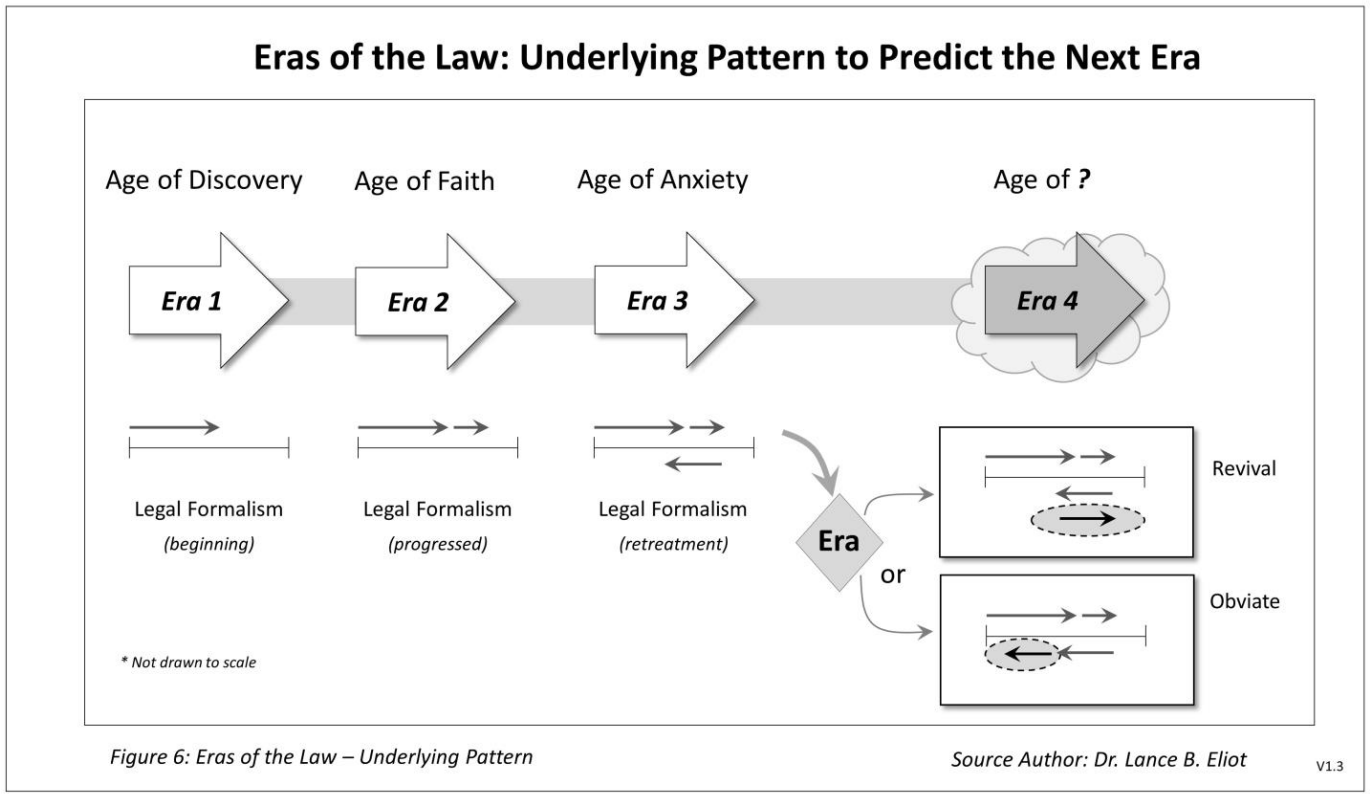